\documentclass[twocolumn,showpacs,preprintnumbers,amsmath,amssymb]{revtex4}

\usepackage{graphicx}
\usepackage{dcolumn}
\usepackage{bm}

\begin{document}


\title { Phase transitions of a tethered surface model with a deficit angle term}

\author{Hiroshi Koibuchi}
 \email{koibuchi@mech.ibaraki-ct.ac.jp}
\author{Zion Sasaki}%
\affiliation{%
Department of Mechanical and Systems Engineering, Ibaraki College of Technology, Nakane 866 Hitachinaka, Ibaraki 312-8508, Japan}%
\author{Keisuke Shinohara}%
\affiliation{%
Technical Support Center, Ibaraki College of Technology, Nakane 866 Hitachinaka, Ibaraki 312-8508, Japan}%


\begin{abstract}
 Nambu-Goto model is investigated by using the canonical Monte Carlo simulations on fixed connectivity surfaces of spherical topology. Three distinct phases are found: crumpled, tubular, and smooth. The crumpled and the tubular phases are smoothly connected, and the tubular and the smooth phases are connected by a discontinuous transition. The surface in the tubular phase forms an oblong and one-dimensional object similar to a one-dimensional linear subspace in the Euclidean three-dimensional space ${\bf R}^3$. This indicates that the rotational symmetry inherent in the model is spontaneously broken in the tubular phase, and it is restored in the smooth and the crumpled phases.

\end{abstract}

\maketitle
\section{Introduction}\label{intro}
A considerable number of studies have been conducted on the phase structure of the elastic membrane model of Polyakov-Kleinert and Helfrich \cite{David-book,Peliti-Leibler,DavidGuitter,David,WHEATER-rev,Wiese-AP,Bowick-PR,Sornette-Ostrowsky,BKS,BK,Kleinert-2,POLYAKOV,Kleinert,HELFRICH}. The Hamiltonian of the model includes not only the Gaussian term but also a bending energy term, which can make the surface smooth. Thus it has been recognized that there are a smooth phase and a crumpled phase in the model of spherical topology. Numerical studies have also been made to understand the phase transitions in the tethered model and in the fluid model on triangulated surfaces \cite{KANTOR-NELSON,KANTOR,GOMPPER-KROLL,WHEATER-cry,BCFTA,BCHHM,ABGFHHM,CATTERALL,KY-IJMPC,KOIB-PLA-2002-2003,Koibuchi-PRE-2003,Koibuchi-PRE-2004-1,AMBJORN-flu}, where the tethered and the fluid models are those defined on fixed connectivity surfaces and on dynamically triangulated surfaces, respectively. 
Except those smooth surfaces and crumpled surfaces, some linear structure \cite{Baillie-Johnston,BEJ,BIJJ,Ferguson-Wheater} and branched polymer surfaces  \cite{Gom-Krol-2,Boal-Rao,Koibuchi-PRE-2003} have also been recognized in the context of surface models.

On the other hand, little is known about a model of tubular surfaces. A tubular surface is considered as one of the basic forms of real physical membranes \cite{Yoshikawa,Hotani}. 

Reviewing references on tubular surfaces, we must recall anisotropic surface models that have been constructed to understand tubular surfaces \cite{Radzihovsky-Toner,Bowick-PR}. A tubular phase is realized in the anisotropic model due to an anisotropic bending modulus. 

However, there has been no study that tried to understand tubular surfaces from an isotropic surface model. Therefore it will be interestng to study an isotropic tethered surface model corresponding to the Nambu-Goto string \cite{Nambu}. The purpose of this study is to understand the phase structure of the tethered surface model of Nambu and Goto with a deficit angle term, which is obtained from the co-ordination dependent term. No bending energy term is included in the Hamiltonian, whereas the deficit angle term serves as a curvature energy smoothing the surface. We will show numerically that the model has three distinct phases; crumpled, tubular, and smooth. The  crumpled and the tubular phases are connected by a higher-order transition, whereas the tubular and the smooth phases are connected by a first-order transition. 

We consider that oblong tubular surfaces can be seen in the Nambu-Goto model. A discretized Hamiltonian of the Nambu-Goto string is given by the sum of the area of triangles, which corresponds to the Gaussian term of the Polyakov-Kleinert model. The area term of the Nambu-Goto model imposes a constraint only on the area of triangles, and hence all the triangles become oblong and form spiky configurations. In fact, it is well known that the partition function of the Nambu-Goto model is not well defined \cite{ADF}. However, it is possible that the partition function of the Nambu-Goto model changes to a well-defined one if some additional term is included in the partition function, as was suggested already in Ref. \cite{ADF}. Then, it is expected that such oblong triangles may form tubular surfaces in such a well-defined model if the additional term tends to modify the spherical surface to a tubular one. 

\section{Model}\label{model}

The area energy $S_1$ is defined by
\begin{equation}
\label{S13-DISC}
S_1=\sum_{\Delta} A_{\Delta},  
\end{equation}
where $A_{\Delta}$ is the area of the triangle ${\Delta}$ in a triangulated surface of spherical topology. This energy $S_1$ of Eq. (\ref{S13-DISC}) is a straightforward discretization of the Nambu-Goto action denoted by $S\!=\!\int d^2x \sqrt{g} $, where $g$ is the determinant of the first fundamental form on the world surface swept out by strings. 

The partition function $Z$ of the Nambu-Goto surface model is defined by 
\begin{eqnarray}
 \label{Z-Nambu-Goto}
Z(\alpha) = \int \prod _{i=1}^N dX_i \exp(-S),\qquad  \nonumber\\
 S(X,{\cal T})=S_1 - \alpha S_3, \quad  
S_3=\sum_i \log (\delta_i/2\pi), 
\end{eqnarray}
where the Hamiltonian $S$ is given by a linear combination of the Gaussian term $S_1$ and the deficit angle term $S_3$, in which $\delta_i$ is the sum of the angles of vertices meeting at the vertex $i$. $S(X,{\cal T})$ denotes that $S$ depends on the embedding $X$ and the triangulation ${\cal T}$ fixed on a uniform lattice, whose construction will be described further in Sec. \ref{MC-Techniques}. The center of the surface is fixed to remove the translational zero mode. $Z(\alpha)$ of Eq. (\ref{Z-Nambu-Goto}) denotes that the model is dependent on the parameter $\alpha$ which is the coefficient of the deficit angle term $S_3$. It should also be noted that the Hamiltonian $S(X,{\cal T})$ is defined only with intrinsic variables of the surface and hence independent of the extrinsic geometries. 

The deficit angle term $S_3$ has a deep connection with the integration measure $dX_i$ \cite{David-NP,BKKM,FN}. 
 We have to remind ourselves that $dX_i$ can be replaced by a weighted measure $dX_iq_i^\alpha $, where $q_i$ is the co-ordination number of the vertex $i$, and $\alpha $ is considered to be $\alpha\!=\!3/2 $. Considering that $q_i$ is a volume weight of the vertex $i$, we think it is possible to assume $\alpha $ as an arbitrary number. Moreover, the co-ordination number $q_i$ can be replaced  by the vertex angle $\delta_i$ according to 
\begin{equation}
 \Pi_i dX_i q_i^\alpha \to \Pi_i dX_i \exp(\alpha \sum_i \log \delta_i). \nonumber
\end{equation}
The constant term $ \sum_i \log 2\pi$ is included to normalize $S_3$ in Eq. (\ref{Z-Nambu-Goto}) so that $S_3\!=\!0$ when $\delta_i\!=\!2\pi$ at every vertex. Thus we have the expression $S_3$ in Eq. (\ref{Z-Nambu-Goto}). Note that $S_3\!=\!0$ not only on the flat surface but also on cylindrical surfaces.

It should also be noted that the deficit angle term $S_3$ is expected to play a nontrivial role in the discrete model which has a finite number of vertices $N$. We expect that $S_3$ is not irrelevant in surface models such as the Nambu-Goto model or the Polyakov-Kleinert model.  In fact, it was reported by Ref. \cite{KOIB-PLA-2002-2003} where the phase transition of the fluid model of Polyakov and Kleinert depends on the coordination dependent term $\sum_i \log q_i$.  

The unit of physical quantities can be explained as follows: The Hamiltonian $S$ in Eq. (\ref{Z-Nambu-Goto}) is obtained from $S\!=\!aS_1 \!-\! \alpha S_3$ by assuming the surface tension coefficient $a$ as $a\!=\!1$. It should be noted that the choice $a\!=\!1$ represents not only a redefinition of $\alpha$ as $\alpha/a$ but also a choice of the unit of length as $\sqrt{kT/a}\!=\!1$, where $T$ is the temperature, and $k$ the Boltzmann constant. Thus the unit of $\alpha$ becomes $kT/a$. The choice of $\sqrt{kT/a}\!=\!1$ for the unit of the length is possible because of the scale invariant property of the partition function in Eq. (\ref{Z-Nambu-Goto}).  

\section{Monte Carlo technique}\label{MC-Techniques}

The triangulated surfaces, on which the model is defined, are uniform in the co-ordination number $q$. On the uniform lattice, the number of vertices $N_q$ of $q\!=\!5$ is $N_5\!=\!12$, and all other vertices are of $q\!=\!6$.  These lattices were obtained by Monte Carlo (MC) simulations with the dynamical triangulation for a model whose Hamiltonian is defined by $S_G\!+\!bS_2\!-\!\alpha S_3(q)$ with sufficiently large $\alpha$, where $S_G\!=\!\sum_i l_i^2 $ is the Gaussian term,  $S_2\!=\!\sum_i (1-\cos \theta_i)$  the bending energy term, and  $S_3(q)\!=\! \sum_i \log q_i$, which is different from $S_3$ in Eq. (\ref{Z-Nambu-Goto}). The well definedness of the uniform lattice must be confirmed, since there are finitely many uniform lattices constructed in this technique for each $N$. We have first confirmed that a fixed connectivity surface model, which is defined by $S_G\!+\!bS_2$, is well defined on such uniform lattices. In fact, the specific heat for the bending energy $S_2$ is independent of the choice of the uniform lattice. 

The canonical Metropolis technique is used to update $X$. The position $X_i$ is moved to a new position $X_i^\prime\!=\!X_i\!+\!{\mit \Delta} X_i$, where ${\mit \Delta} X_i$ is randomly chosen in a small sphere. $X_i^\prime$ is accepted with the probability ${\rm Min}[1,\exp\left(-{\mit \Delta}S\right) ]$, where ${\mit \Delta}S\!=\!S({\rm new})\!-\!S({\rm old})$. The radius of the small sphere for ${\mit \Delta} X_i$ is chosen at the beginning of the simulations to maintain 35 $\sim$ 60 $\%$ of acceptance rate; almost all MC simulations are done on about 50 $\%$ of acceptance rate.

A lower bound $10^{-6}A_0$ is imposed on the area of triangles in the update of $X$, where $A_0$ is the mean area of the triangles computed at every 250 MCSs (Monte Carlo sweeps) and $A_0$ is constant due to the relation $S_1/N\!=\!1.5$. However, the areas are almost free from such constraint, because the areas of almost all triangles are larger than $10^{-6}A_0$ throughout the MC simulations. No constraint is imposed on the bond length.

\section{Results}\label{Results}

\begin{figure}[htb]
\includegraphics[width=8.5cm]{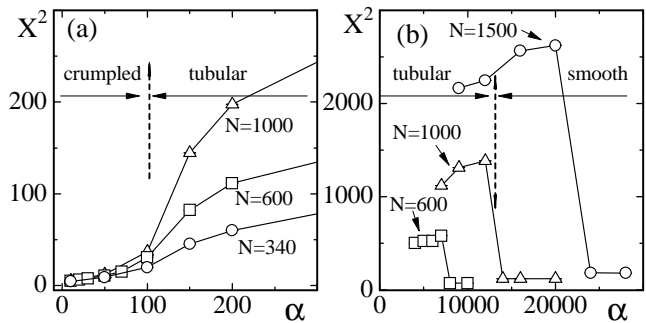}
\caption{(a) $X^2$ obtained at relatively small $\alpha$ in the vicinity of the boundary between the crumpled and the tubular phases,  (b) $X^2$ obtained at relatively large $\alpha$ in the vicinity of the boundary between the tubular and the smooth phases. The unit of $X^2$ and $\alpha$  is $kT/a$, where $a$ is the surface tension.  } 
\label{fig-1}
\end{figure}
We first show in Figs. \ref{fig-1}(a) and (b) the mean square size $X^2$ defined by
\begin{equation}
\label{Spec-Heat-S3}
X^2= {1\over N} \sum _i \left( X_i-\bar X \right)^2,\quad \bar X = {1\over N} \sum_i X_i,
\end{equation}
where $\bar X$ is the center of the surface. $X^2$ in Fig. \ref{fig-1}(a) represents that the size of surfaces continuously increases with increasing $\alpha$, and also represents that the shape of the surfaces rapidly changes at $\alpha\simeq 100$. On the other hand, $X^2$ shown in Fig. \ref{fig-1}(b) clearly represents some discontinuous transition, where $X^2$ abruptly changes. The dashed lines drawn vertically on the data of the $N\!=\!1000$ surface in both of the figures represent the phase boundaries, on which we focused our attention in this paper.  

The convergence speed of MC is very low in the tubular phase close to the smooth phase. The total number of MCS at $\alpha\!=\!20000$ on the $N\!=\!1500$ surface is about $2\times 10^9$, where $1.5\times 10^9$ MCSs were discarded for the thermalization. This is the reason why we use surfaces of size up to $N\!=\!1500$. The reason of the low convergence speed seems due to a straight-line structure of the surface, which will be shown below. Since the vertices can move only along the line, the surface deforms very slowly. On the contrary, the convergence both in the smooth phase and in the crumpled phase is relatively faster than that in the tubular phase. In the simulation the expected relation $S_1/N\!=\!3/2$ is satisfied in the configurations reached after the thermalization at every $\alpha$. 

\begin{figure}[htb]
\includegraphics[width=8.5cm]{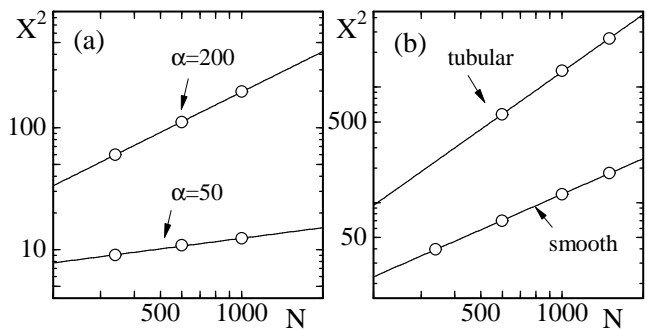}
\caption{(a) Log-log plot of $X^2$ vs $N$ at $\alpha\!=\!50$ (crumpled phase), and $\alpha\!=\!200$ (tubular phase),  and (b) those obtained at $\alpha$ close to the phase boundary of the discontinuous transition. The unit of $X^2$ is $kT/a$.  }
\label{fig-2}
\end{figure}
Figure \ref{fig-2}(a)  shows log-log plots of $X^2$ against $N$ obtained at $\alpha\!=\!50$ and $\alpha\!=\!200$.  Plots of $X^2$ against $N$ in Fig. \ref{fig-2}(b) denoted by {\it tubular} ({\it smooth}) were obtained below (above) the discontinuous transition point in each $N$ as shown previously in Fig. \ref{fig-1}(b). The straight lines plotted in Figs. \ref{fig-2}(a) and (b) are those fitted by
\begin{equation}
\label{X2-scale}
X^2 \propto N^{2/H},
\end{equation}
where $H$ is the Hausdorff dimension. From the slope of the plotted lines, we have
\begin{equation}
\label{Hausd}
H_{50}=7.24\pm0.48, \quad H_{\rm smo}=1.93\pm0.01,
\end{equation}
where $H_{50}$ and $H_{\rm smo}$ were obtained from the data denoted by $\alpha\!=\!50$ in Fig. \ref{fig-2}(a) and those by {\it smooth} in Fig. \ref{fig-2}(b), respectively. Those results are in agreement with our expectation. In fact, $H$ is expected to be very large in the crumpled phase, and it is also expected to be $H\!=\!2$ in the smooth phase. Moreover, we have $H_{200}\!=\!1.80\pm0.02$ and $H_{\rm tub}\!=\!1.22\pm0.03$, which were obtained from the data denoted by $\alpha\!=\!200$ in Fig. \ref{fig-2}(a) and those by {\it tubular} in Fig. \ref{fig-2}(b), respectively. $H_{200}$ and $H_{\rm tub}$ slightly deviate from $H\!=\!2$ which is confirmed in the case of branched polymer surfaces \cite{Koibuchi-PRE-2003}, where surfaces randomly stretches and hence are rotationally symmetric. 

\begin{figure}[htb]
\includegraphics[width=8.5cm]{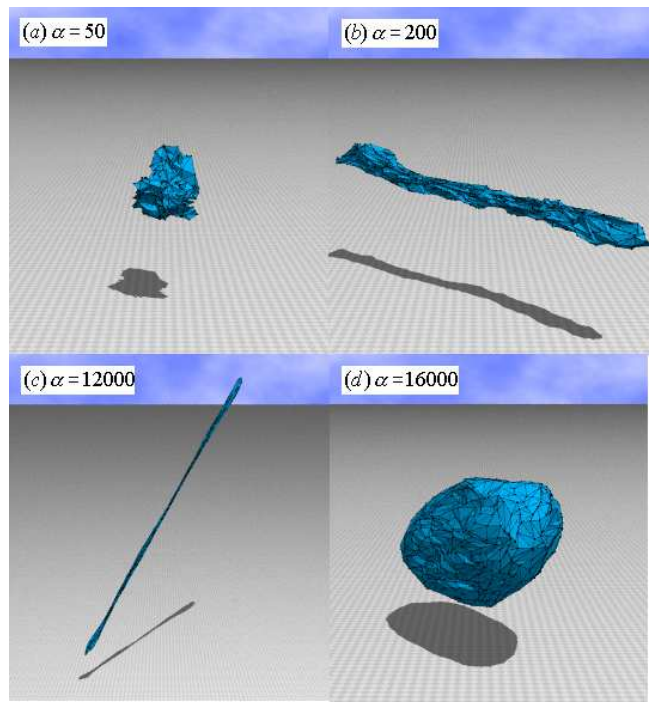}
\caption{Snapshots of $N\!=\!1000$ surfaces obtained at (a) $\alpha\!=\!50$ (crumpled), (b) $\alpha\!=\!200$ (tubular), (c) $\alpha\!=\!12000$ (tubular), and (d) $\alpha\!=\!16000$ (smooth).  Surfaces in (a), (b), and (d) are drawn in the same scale, which is different from that in (c).}
\label{fig-3}
\end{figure}
Snapshots of $N\!=\!1000$ surfaces are shown in Figs. \ref{fig-3}(a)--(d) obtained at $\alpha\!=\!50$,  $\alpha\!=\!200$,  $\alpha\!=\!12000$,  and $\alpha\!=\!16000$. Figures \ref{fig-3}(a), (b), and (d) were drawn in the same scale, which is different from that in Fig. \ref{fig-3}(c). The axis direction of the surface in Fig. \ref{fig-3}(c) is spontaneously chosen. The direction of the axis remains almost unchanged throughout the MC simulation. Thus we find no tubular surface bending in the tubular phase for $N\!\leq\!1500$. The straight-line structure shown in Fig. \ref{fig-3}(c) is expected to survive even at a sufficiently large $N$. The reason is because both $S_3$ and oblong triangles tend to straighten the surface. 

\begin{figure}[htb]
\includegraphics[width=8.5cm]{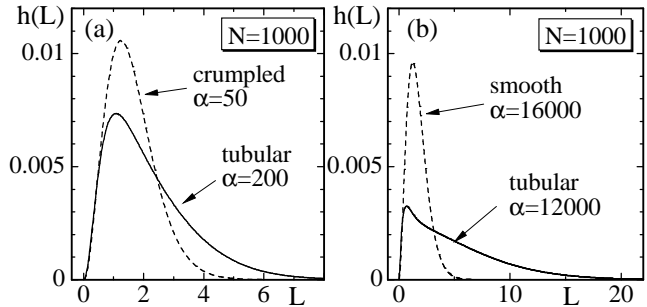}
\caption{Normalized distribution $h(L)$ of the bond length $L$ obtained at (a) $\alpha\!=\!50$ (crumpled), and $\alpha\!=\!200$ (tubular), and at (b) $\alpha\!=\!12000$ (tubular), and $\alpha\!=\!16000$ (smooth), on $N\!=\!1000$ surfaces. The unit of $L$ is $\sqrt{kT/a}$.  }
\label{fig-4}
\end{figure}
Figures \ref{fig-4}(a) and (b) are normalized distribution $h(L)$ of the bond length $L$ sampled  at every 500 MCSs in the final $2\!\times\!10^7$ MCSs on $N\!=\!1000$ surfaces. The normalization of $h(L)$ is  given by $\sum_i h(L)/{\mit \Delta}L\!=\!1$, where ${\mit \Delta} L\!=\!0.02$, and the sum $\sum_i$ runs over all bonds and hence $\sum_i 1$ becomes identical with $N_B$ the total number of bonds. The dashed and solid curves denoted by {\it crumpled} and {\it tubular} in Fig. \ref{fig-4}(a) were obtained at $\alpha\!=\!50$ and $\alpha\!=\!200$ respectively, and those denoted by {\it smooth} and {\it tubular} in Fig. \ref{fig-4}(b) were obtained at $\alpha\!=\!16000$ and $\alpha\!=\!12000$, respectively. 

 We note that $h(L)$ obtained on surfaces of size other than $N\!=\!1000$ are exactly identical with $h(L)$ in Fig. \ref{fig-4}(a) if $\alpha$ is identical with each other.  Moreover, $h(L)$ obtained at the smooth phase, denoted by {\it smooth}  in Fig. \ref{fig-4}(b), is independent of both $N$ and $\alpha$ in the smooth phase. While $h(L)$ is dependent on $\alpha$ in the tubular phase close to the crumpled phase, it is almost independent of both $N$ and $\alpha$  in the tubular phase close to the smooth phase. The fluctuations of the surface size in the tubular phase close to the smooth phase are relatively larger than those in the tubular phase close to the crumpled phase.

It should be noted that the distribution $h(A)$ of the area $A$ of the triangles, which is not presented in a figure, is universal in a sense that $h(A)$ is independent not only of $N$ but also of $\alpha$. $h(A)$ is not influenced even by the discontinuous transition. In fact, $h(A)$ obtained in the tubular phase is exactly identical not only with that in the smooth phase but also with that in the crumpled phase.

\begin{figure}[htb]
\includegraphics[width=8.5cm]{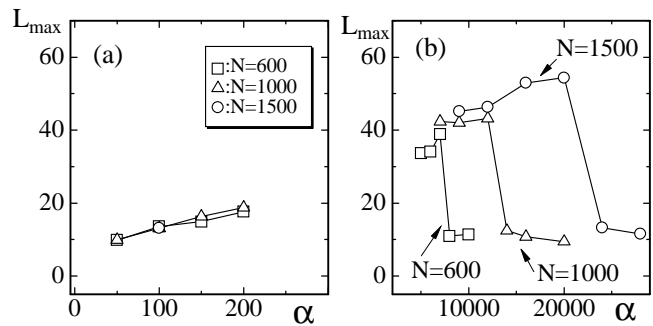}
\caption{Maximum bond length $L_{\rm max}$  at (a) relatively small $\alpha$, and at (b)  relatively large $\alpha$. The unit of $L_{\rm max}$ is $\sqrt{kT/a}$, and that of $\alpha$  is $kT/a$   }
\label{fig-5}
\end{figure}
It must be checked that the size of triangles is negligible compared to the size of surfaces at sufficiently large $N$. In order to see that the maximum bond length $L_{\rm max}$ is considerably smaller than the size of surfaces in the tubular phase, we plot in Figs. \ref{fig-5}(a) and (b) $L_{\rm max}$ obtained in the final $2\!\times\!10^7$ MCSs on each surface. We find in Fig. \ref{fig-5}(a) that $L_{\rm max}$ continuously increases with $\alpha$ at the boundary between the crumpled and the tubular phases, and that $L_{\rm max}$ at each $\alpha$ is almost independent of $N$.  In the tubular phase at $\alpha\!=\!200$, $L_{\rm max}$ is smaller than the surface length $L_s$: 
\begin{eqnarray}
\label{Lsize}
L_s\simeq 41 \quad (N=600,\;\alpha=200), \nonumber \\
L_s\simeq 56 \quad (N=1000,\;\alpha=200).
\end{eqnarray}
The length $L_s$ were obtained by $L_s\!=\!\sqrt{\langle L_s^2\rangle}$, where  $\langle L_s^2\rangle$ was obtained in the tubular phase close to the crumpled phase. While $L_{\rm max}$ in Fig. \ref{fig-5}(a) is almost independent of $N$, $L_s$ increases with $N$ at $\alpha\!=\!200$ as shown in Eq. (\ref{Lsize}). Hence it is expected that $L_{\rm max}/L_s\!\to\! 0$ in the limit $N\!\to\! \infty$ at least in the tubular phase close to the crumpled phase.

 Figure \ref{fig-5}(b) shows that $L_{\rm max}$ in the tubular phase gradually increases as $N$ increases. However, we find that $L_{\rm max}$ is considerably smaller than $L_s$ in the tubular phase close to the smooth phase. In fact, we have $L_{\rm max}\!\simeq \! 43$ for $N\!=\!1000$ and $L_{\rm max}\!\simeq \! 54$ for $N\!=\!1500$, which are smaller than the length of the surfaces shown below:
\begin{eqnarray}
L_s\simeq 122 \quad (N=1000,\;\alpha=12000),\nonumber \\
L_s\simeq 172 \quad ( N=1500,\;\alpha=20000).\nonumber
\end{eqnarray}
Therefore it is also expected that $L_{\rm max}/L_s\!\to\! 0$ in the limit $N\!\to\! \infty$ everywhere in the tubular phase. 

\begin{figure}[htb]
\includegraphics[width=8.5cm]{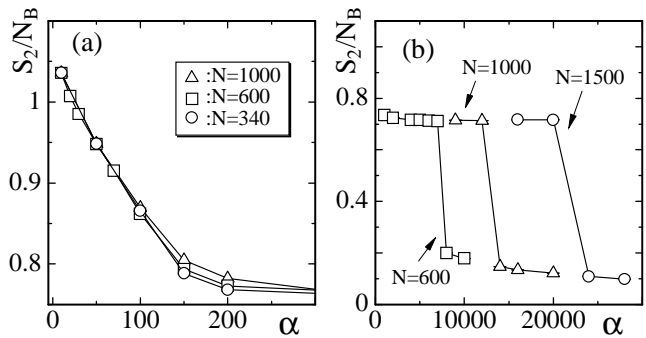}
\caption{The bending energy $S_2/N_B\!=\!\sum_i(1-\cos \theta_i)/N_B$ at (a) relatively small $\alpha$,  and at (b) relatively large $\alpha$. $N_B$ is the total number of bond. The unit of $\alpha$  is $kT/a$.  }
\label{fig-6}
\end{figure}
Finally, we plot in Figs.  \ref{fig-6}(a) and (b) the bending energy $S_2\!=\!\sum_i(1-\cos \theta_i)$, which reflects a smoothness of surfaces and it is not included in the Hamiltonian. While $S_2/N_B$ continuously changes against $\alpha$ in Fig. \ref{fig-6}(a), it is clearly discontinuous in Fig. \ref{fig-6}(b). These results indicate that the tubular phase is smoothly connected to the crumpled phase and discontinuously connected to the smooth phase. 
The higher-order nature of the transition between the tubular and the crumpled phases has also been seen in the specific heat $C_{S_3}\!=\!(\alpha^2/N)\left( \langle S_3^2 \rangle \!-\!\langle S_3 \rangle ^2 \right)$. In fact, although $C_{S_3}$ has a peak at $\alpha\!\simeq\!100$, there was no growth of the peak with increasing $N$.   

The bending energy $S_2/N_B$ is not an order parameter regarding to the tubular phase because it is also nonzero in the crumpled phase. However, $S_2/N_B$ plays a role of order parameter, as can be seen in Fig. \ref{fig-6}(b), as far as we confine ourselves to the transition between the tubular phase and the smooth phase.   

\section{Summary and conclusion}\label{Conclusions}

We have investigated the phase structure of a tethered surface model of Nambu-Goto embedded in ${\bf R}^3$, and found that there are three distinct phases: smooth, tubular, and crumpled. Moreover, the model undergoes a first-order transition between the smooth and the tubular phases, and a higher-order transition between the tubular and the crumpled phases. The surface forms an oblong and one-dimensional object in the tubular phase. It is remarkable that the rotational symmetry or the symmetry of isotropy inherent in the model is spontaneously broken in the tubular phase.

 An important point to emphasize is that both terms area $S_1$ and deficit angle $S_3$ are the cause of such variety of phases. Moreover, it is quite likely that the straight-line structure in the tubular phase survives even at sufficiently large $N$, because not only the oblong triangles but also the deficit angle term can make the surface tubular. 

 Further numerical studies on the fluid model and on the model with extrinsic curvature would give us hints to clarify the phase diagram of the Nambu-Goto surface model with the deficit angle term.

\begin{acknowledgments}
This work was supported in part by a Grant-in-Aid for Scientific Research Grant No. 15560160. H.K. thanks N. Kusano, A. Nidaira, and K. Suzuki for their invaluable help in the early stage of the work.  H.K. would like to thank T. Hamada for discussions on real tubular membranes during the YITP workshop YITP-W-03-06 on Soft Matter Physics 2003. H.K. would like to thank L. A. Guzman for reading the manuscript. 
\end{acknowledgments}



\end{document}